# EDUCATION-CENTERED CRITICAL POLICY ANALYSIS OF AI: GHANA'S AI STRATEGY AS A CASE

Matthew Nyaaba[1], Vida Awinime Bugri [2], Eric Kojo Majialuwe[3,5], Bismark Nyaaba Akanzire[4], Ibrahim Nantomah[5], Felicia Boateng[6], Patrick Kyeremeh[7], Benjamin Quarshie[8], Ellen Kwarteng[9], and Macharious Nabang[10]

[1] *University of Georgia, USA.*
[2] *Université de Guyane, France.*
[3] *University of Manitoba, Canada.*
[4] *Gambaga College of Education. Gambaga, Ghana.*
[5] *University for Development Studies, Ghana.*
[6] *The Open University, UK.*
[7] *St. Joseph's College of Education, Bechem, Ghana.*
[8] *Mampong Technical College of Education, Ghana.*
[9] *St. Joseph's College of Education, Bechem, Ghana.*
[10] *Bagabaga College of Education, Tamale, Ghana.*

ABSTRACT

National AI strategies are increasingly analyzed for their attention to governance, workforce development, innovation, and competitiveness, but less is known about how they frame education as a sector with its own pedagogical, cultural, ethical, and implementation demands. This study addresses this gap by developing and applying an Education-Centered AI Policy Framework to critically analyze Ghana's *National Artificial Intelligence Strategy, 2025–2035*. Using critical qualitative policy document analysis, we deductively examined the strategy through the six components of the Education-Centered AI Policy Framework. The findings show that Ghana's strategy is ambitious and timely, especially in its emphasis on AI literacy, youth skills, TVET, workforce readiness, rural outreach, local language data, inclusion, and responsible AI governance. However, the education agenda is stronger on national AI readiness than on school-level implementation. Teacher agency, pre-service teacher education, curriculum progression, assessment guidance, AI disclosure, multilingual pedagogy, culturally responsive AI use, child-centered safeguards, and participatory governance remain underdeveloped. The analysis also identifies document-level concerns, including apparent AI-styled visual content without visible disclosure and a mismatch between the strategy's vision and mission figure and the related textual explanation, raising questions about the document's own transparency, coherence, and modeling of responsible AI communication. We argue that Ghana needs a sector-specific, education-centered AI policy and implementation pathway that includes key education stakeholders and connects workforce readiness with teacher preparation, curriculum reform, assessment redesign, learner protection, infrastructure, local language instruction, culturally responsive pedagogy, locally responsive AI tools, and participatory governance.

## Introduction

Artificial intelligence (AI) is becoming a major policy concern for governments seeking to strengthen education, workforce development, innovation, public service delivery, and economic competitiveness. In response, many countries, including Canada, South Korea, Egypt, and China, have developed national AI strategies to guide adoption, investment, regulation, and institutional coordination across sectors (Radu, 2021; Schiff, 2022; Shi, 2025). These strategies reveal how governments imagine the future of learning, labour, governance, and national development. For countries in sub-Saharan Africa, this policy movement is especially important because the region has historically faced structural barriers in responding to major technological shifts (Amankwah-Amoah, 2019). Amankwah-Amoah (2019) notes that throughout the second half of the twentieth century, many sub-Saharan African countries, particularly in West Africa, "remained largely stagnant in terms of innovations" while other regions advanced more rapidly (p. 912).

Against this backdrop, *Ghana's National Artificial Intelligence Strategy, 2025-2035*, marks an important step in this global policy movement. The strategy presents a ten-year vision to position Ghana as a leading AI hub in Africa by 2035 and identifies education as a first and major pillar of national AI transformation (Ministry of Communication, Digital Technology and Innovations [MoCDTI], 2025). This emphasis is significant because Ghana's AI future will depend not only on infrastructure and investment, but also on how schools, teachers, learners, communities, and educational institutions are prepared to engage with AI in meaningful and responsible ways (Funa & Gabay, 2025; Reimers et al., 2026).

However, identifying education as a policy priority does not necessarily mean that education is fully understood as a complex social, cultural, pedagogical, and ethical system. Existing studies show that national AI strategies often frame education mainly as a pathway for producing AI talent and preparing future workers. Schiff's (2022) analysis of 24 national AI strategies found that many policies emphasized "education for AI," especially the preparation of technical experts, AI professionals, and digitally skilled workers. Far less attention was given to "AI for education," including how AI may reshape teaching, learning, assessment, learner protection, equity, and classroom practice. Similarly, Shi's (2025) analysis of 50 national AI strategies found that countries frequently prioritized AI workforce competence through higher education reform, lifelong learning, reskilling, and labour-market training. These studies suggest a persistent policy gap, education is often treated as a support system for national AI readiness rather than as a sector requiring its own pedagogical, ethical, cultural, and governance frameworks.

Ghanaian education is shaped by multilingual classrooms, cultural diversity, rural and urban inequalities, uneven digital access, teacher preparation challenges, and ongoing curriculum reforms Recent studies on AI and education in the Global South reinforce this concern. Reimers et al. (2026) argue that the value of AI in education depends strongly on existing educational conditions, including infrastructure, teacher preparation, institutional capacity, pedagogical traditions, and local cultural expectations. Their analysis cautions against importing AI tools or policy models without examining how they fit the realities of low-resource and culturally diverse education systems. In Ghana, studies on culturally responsive AI further show that educational AI tools become more meaningful when they reflect local languages, indigenous knowledge systems, curriculum expectations, and learners' lived experiences (Nyaaba & Zhai, 2025). These findings suggest that responsible AI policy in education must move beyond technical readiness to address the cultural and pedagogical conditions under which AI will be used.

Research on national AI governance shows that policy processes are often shaped by government agencies, technical experts, academia, and private-sector actors, while affected communities and end-users remain less visible (Radu, 2021). This imbalance is particularly consequential in education, where AI policy directly affects teachers, students, parents, school leaders, teacher educators, disability advocates, language communities, rural schools, and local communities. Their participation matters because responsible AI integration in education must address practical and ethical issues such as "Academic Integrity, Transparency in AI Use, Human Oversight, Data Privacy, and Regulation and Monitoring" (Funa & Gabay, 2025, p. 5). In Ghana, these concerns intersect with broader educational realities, including AI literacy, inclusion, pedagogy, assessment, learner protection, local languages, cultural responsiveness, rural access, and the representation of diverse subcultures (Funa & Gabay, 2025; Nyaaba & Zhai, 2025). Guided by critical policy analysis, this study examines Ghana's *National Artificial Intelligence Strategy, 2025–2035* through an education-sector lens (Diem et al., 2014). It asks how education is framed, whose voices are visible, what educational issues are emphasized or overlooked, and whether the strategy provides sufficient guidance for responsible AI use in schools and teacher education.

## Literature Review

### National AI Strategies and Education

National AI strategies have become important policy instruments through which governments define priorities for innovation, workforce development, regulation, public service delivery, and national competitiveness. Education is often central to these strategies because AI readiness depends on human capacity, technical expertise, digital skills, and public understanding. However, studies show that national AI strategies do not always treat education as a full pedagogical, cultural, and ethical system. Radu (2021) found that national AI policy processes often privilege government, industry, academia, and technical experts, while broader public and end-user participation remains less visible.

The education gap is clearer in studies that examine how national AI strategies frame learning. Schiff (2022) found that many strategies emphasize "Education for AI," especially the preparation of AI experts, technical workers, and future labour-market participants. By contrast, "AI for Education," including classroom practice, assessment, learner protection, teacher agency, and equity, receives less attention. Shi's (2025) analysis of 50 national AI strategies similarly found that workforce preparation, higher education reform, reskilling, and lifelong learning dominate national education priorities. These findings suggest that education is often positioned as a pipeline for national AI readiness rather than as a sector requiring its own implementation, ethical, and pedagogical frameworks.

This distinction matters for Ghana because AI policy will be enacted within classrooms shaped by multilingualism, urban-rural disparities, examination pressures, teacher preparation needs, uneven digital access, and ongoing curriculum reform. If Ghana's AI strategy frames education mainly as talent development, then issues such as assessment integrity, learner data protection, teacher professional judgment, local language inclusion, and culturally responsive pedagogy may remain secondary. This paper, therefore, examines not only whether education appears in the strategy, but also how education is imagined and whose educational realities are made visible.

**AI in Education**

Research on AI in education points to two connected policy needs. The first is AI literacy. Students need to understand what AI is, how it works, where it fails, and how it should be used responsibly. Touretzky et al. (2019) proposed age-appropriate "big ideas" for K-12 AI learning, while Casal-Otero et al. (2023) found that K-12 AI literacy research has grown around learning experiences and theoretical models but still needs stronger curriculum continuity and evaluation. Chee et al. (2024) extend this discussion by showing that AI literacy should include data and algorithm literacy, ethics, problem solving, communication, collaboration, career awareness, content creation, and affective competence. These studies indicate that AI literacy should not be reduced to coding or robotics alone.

The second need concerns how AI changes teaching and learning. Generative AI can support tutoring, feedback, content generation, lesson planning, differentiated instruction, and administrative work. Kasneci et al. (2023) and Yan et al. (2024) show that these tools may expand personalized support and learning resources, but they also create risks related to misinformation, bias, over-reliance, academic integrity, and weak transparency. Yan et al. (2024) therefore argue that education systems must support both learning with AI and learning about AI. This finding is important for policy because AI literacy and AI-supported pedagogy cannot be treated as separate reforms.

Assessment has become one of the strongest areas of current AI-in-education research. Xia et al. (2024) found that AI affects assessment at the student, teacher, and institutional levels, especially through feedback, academic integrity, task design, and assessment literacy. Perkins et al. (2023) propose that institutions should make different levels of AI use explicit and align them with learning outcomes rather than treating all AI use as misconduct. These studies show a shift from detecting AI use to redesigning assessment around reasoning, process, reflection, transparency, and human judgment. However, much of this evidence comes from higher education, leaving an important gap in school-level AI assessment policy, especially for younger learners and examination-driven systems.

These issues also connect directly to teacher agency. Teachers are not only implementers of AI policy; they are professional decision-makers who must evaluate AI outputs, adapt materials to context, protect learners, and decide when human judgment should override automated suggestions. Mishra et al. (2023) argue that generative AI requires expanded technological, pedagogical, and contextual knowledge, while Zhai (2024) frames teacher agency as a developmental movement from observer to adopter, collaborator, and innovator. Studies on teacher education also show that pre-service teachers often use AI for lesson planning and resources, but need stronger preparation in ethics, inclusion, facilitation, and critical evaluation (Vaughan & Wah, 2026; Yadav et al., 2025). For Ghana, this evidence points to the need for a clear teacher AI competency pathway rather than broad training commitments.

### Responsible AI, Student Data, and Educational Ethics

Responsible AI in education requires more than general ethical language. It requires rules and practices for privacy, transparency, accountability, safety, fairness, disclosure, consent, and human oversight. Nguyen et al. (2022) identify governance, transparency, privacy, security, inclusiveness, and human-centeredness as key ethical principles for AI in education. These principles are particularly important in schools because educational AI systems may collect sensitive learner data, influence assessment, generate profiles of students, and shape learning opportunities.

Studies on algorithmic fairness show why these safeguards are necessary. Holstein and Doroudi (2021) warn that AI can reproduce educational inequalities even when equity is a stated goal, while Kizilcec and Lee (2020) show that fairness concerns can appear across measurement, model learning, and educational action. In practice, this means that biased data or poorly evaluated systems can affect how learners are assessed, supported, recommended, or excluded. Lee et al. (2024) further show that educational AI governance must address risks such as opacity, discrimination, privacy violations, and weak accountability.

For national policy, the key issue is implementation detail. A strategy may mention ethics, privacy, and responsible AI, but schools still need practical guidance on student data ownership, consent, data storage, cross-border transfer, algorithmic profiling, AI disclosure, human review, and academic integrity. This is especially important in Ghana, where AI integration will involve children, teachers, public institutions, private vendors, and uneven digital infrastructure. Responsible AI in education therefore requires sector-specific safeguards, not only broad national principles.

### Culturally Responsive AI

Language and culture are central to AI policy in Ghana because they shape how learners understand, communicate, and connect school knowledge to community life. Many AI systems are trained mainly on high-resource languages, especially English, which can marginalize African languages and cultural knowledge systems. Ożegalska-Łukasik and Łukasik (2023) show how this language imbalance can limit inclusion, while Issaka et al. (2024) identify the scarcity of digitized corpora, annotated datasets, and speech resources for Ghanaian languages as a major limitation for natural language processing.

This issue is educational, not only technical. Owu-Ewie and Eshun (2019) show that Ghana's multilingual classrooms are shaped by movement between home languages and English-medium instruction. When learners are expected to engage abstract ideas mainly through unfamiliar language structures, comprehension and participation may be affected. Recent initiatives show progress. Gyamfi et al. (2026) report parallel sentence resources for Ghanaian languages, while Wiafe et al. (2025) describe large-scale speech datasets across languages such as Akan, Ewe, Dagbani, Dagaare, and Ikposo. Still, these resources remain limited for full educational AI deployment because classroom language use includes dialects, cultural meanings, local examples, and pragmatic forms of expression.

Culturally responsive AI addresses this gap by asking whether AI tools reflect learners' linguistic, cultural, and educational realities. In Ghanaian education, this includes local language support, indigenous knowledge, curriculum-aligned examples, culturally meaningful scenarios, and rural-urban differences. Studies on culturally responsive AI tools in Ghana suggest that AI systems designed with local languages, indigenous knowledge, and contextual examples can produce more relevant educational outputs than generic prompting approaches (Nyaaba & Zhai, 2025; Nyaaba et al., 2025). This finding is important for the present study because it positions cultural responsiveness not as an optional addition, but as a condition for meaningful AI implementation in Ghanaian education.

### Stakeholder Participation in AI Policy

Stakeholder participation is central to AI governance because AI policies affect groups differently depending on their location, resources, language, role, and level of power. Radu (2021) found that national AI policy processes often involve government, academia, industry, and technical actors more visibly than civil society, end-users, and marginalized communities. In education, this imbalance matters because teachers, students, parents, school leaders, teacher educators, rural communities, disability advocates, and language communities are the people who experience AI policy most directly.

Participatory governance research shows that meaningful involvement requires more than consultation. Arnstein (1969) distinguishes tokenistic participation from forms of participation that allow affected groups to influence decisions. Gaber (2020) similarly argues that participation should enable marginalized actors to shape policy outcomes rather than simply respond to decisions already made. The OECD (2019) links inclusive participation to trustworthy AI, while White and Langenheim (2021) show that participation must be designed in ways that give communities real influence. These studies are important because educational AI policy can easily become expert-led while classroom actors remain peripheral.

In education, participation is not only a democratic value; it is also an implementation condition. Teachers understand classroom realities, assessment pressures, learner needs, and professional constraints. Students can identify concerns about surveillance, fairness, privacy, consent, and over-reliance, as Burriss et al. (2024) found in studies involving middle and high school students. Parents and communities also shape trust, cultural legitimacy, and acceptance of AI-supported learning. For Ghana, this means that education-sector AI governance should involve those who understand rural schooling, local languages, disability inclusion, teacher preparation, community values, and learner protection.

### Existing AI Policy and Education Frameworks

Existing AI policy frameworks often organize AI governance around national competitiveness, regulation, risk management, and technology adoption. For instance, Wang et al. (2025) found that China emphasizes research and application, the European Union emphasizes social impact, and the United States emphasizes government role. While such comparative work is useful for understanding national AI governance, it gives limited attention to the educational questions that emerge when AI enters classrooms, teacher preparation, assessment, and learner protection. Higher education frameworks offer more education-specific guidance, but they remain largely institution-centered. Li et al. (2024) proposed a university framework for generative AI governance based on perceived usefulness, perceived risk, facilitating conditions, social influence, and self-efficacy. Chan (2023) similarly developed an AI education policy framework organized around pedagogical, governance, and operational dimensions. However, these frameworks remain largely focused on university-level adoption and do not fully address school-level implementation, teacher agency, cultural responsiveness, and community participation.

K-12 and AI literacy frameworks extend the discussion but still leave important policy gaps. For instance, Eutsler et al. (2026) show that school district AI policies remain emergent and often focus on academic integrity, responsible use, teacher guidance, data privacy, and educational support. AI literacy and STEM-oriented frameworks further emphasize knowledge, evaluation, contextualization, ethics, assessment redesign, and learner agency (Allen & Kendeou, 2024; Leon et al., 2025). In all, these studies offer useful insights, but they do not fully connect national AI policy purpose with teacher agency, curriculum, pedagogy, assessment, learner protection, language, culture, and participatory implementation. This gap supports the need for an *Education-Centered AI Policy and Participatory Governance Framework* that treats AI in education as a pedagogical, ethical, cultural, and governance issue.

### Toward an Education-Centered AI Policy Framework

This study is grounded in critical policy analysis, which treats policy as more than a neutral technical response to public problems. Diem et al. (2014) argue that critical policy analysis asks whose interests and assumptions shape policy, including "who is behind the policy" and "whose voice is being privileged" (p. 1077). They also emphasize the need to examine "what's missing" and "what are the silences" in policy texts (p. 1077). This orientation is important for the present study because Ghana's *National Artificial Intelligence Strategy, 2025–2035* does not only describe AI priorities; it also constructs meanings of education, AI readiness, inclusion, governance, and national development. Building on this critical policy orientation and the reviewed literature, this paper proposes an Education-Centered AI Policy Framework for analyzing national AI strategies as education policy documents, not only as technology, innovation, or workforce plans. As shown in Figure 1, the framework places *policy purpose* at the center and asks how education is framed within AI policy. Around this center are four education-centered dimensions: *teacher agency and professional learning; curriculum, pedagogy, and assessment; language, culture, and contextual responsiveness; and responsible AI and learner protection*. These dimensions are surrounded by *participation and implementation governance*, which focuses on who is involved in shaping, implementing, monitoring, and revising AI policy in education.

At the center of the framework, *policy purpose* examines whether education is framed mainly as workforce preparation or as a broader site of pedagogical, cultural, ethical, and institutional transformation. This focus draws from Diem et al. 's (2014) attention to policy construction by asking how AI readiness is defined, what kind of educational future is imagined, and which purposes of education are privileged. Schiff (2022) shows that national AI strategies often emphasize "education for AI," while Shi (2025) similarly finds that workforce preparation, reskilling, and national competitiveness frequently dominate national AI education agendas. The first dimension, *teacher agency and professional learning*, considers how teachers, teacher education, and professional development are positioned in AI policy. Diem et al.'s (2014) concern with how policy "rolls out" to those affected by it is especially useful here, since teachers are the actors who translate national AI policy into classroom practice. Mishra et al. (2023) argue that generative AI requires expanded technological, pedagogical, and contextual knowledge, while Zhai (2024) positions teacher agency as central to meaningful AI integration. Teacher professional learning should therefore move beyond tool training to support ethical judgment, bias detection, learner protection, contextual adaptation, and human oversight.

The second dimension, *curriculum, pedagogy, and assessment*, addresses how AI literacy, classroom practice, and evaluation are defined. Informed by Diem et al.'s (2014) call to unpack policy assumptions, this dimension asks whether AI education is treated as coding and productivity alone or whether it also includes learning, authorship, assessment, and academic integrity. Casal-Otero et al. (2023) show the need for stronger K-12 AI literacy progression, and Chee et al. (2025) frame AI literacy as including data literacy, algorithmic understanding, ethics, communication, problem solving, and affective competence. Perkins et al. (2024) further show that assessment policy must clarify acceptable levels of AI use, while Xia et al. (2024) demonstrate that generative AI is reshaping feedback, task design, transparency, and human judgment. The third dimension, *language, culture, and contextual responsiveness*, asks whether AI policy attends to local languages, indigenous knowledge, cultural identities, and Ghanaian educational realities. This dimension extends Diem et al.'s (2014) concern with policy silences by examining which cultural, linguistic, rural, and community-based knowledge systems are present, marginal, or absent in national AI policy. Owu-Ewie and Eshun (2019) show that Ghanaian classrooms are multilingual spaces where language shapes participation and learning. Issaka et al. (2024) identify gaps in Ghanaian natural language processing resources, while Wiafe et al. (2025) show the importance of developing speech datasets for low-resource Ghanaian languages. Nyaaba and Zhai (2025) further show that culturally responsive AI becomes more educationally meaningful when it draws on local languages, indigenous knowledge, and curriculum-aligned cultural examples.

The fourth dimension, *responsible AI and learner protection*, focuses on whether AI policy provides safeguards for privacy, consent, transparency, fairness, disclosure, accountability, and human oversight (UNESCO, 2021). This part of the framework reflects Diem et al.'s (2014) concern with who is affected by policy and whether policy does what it claims to do. Nguyen et al. (2022) argue that responsible AI in education requires attention to transparency, governance, privacy, fairness, and accountability. Holstein and Doroudi (2021) warn that AI can reproduce educational inequalities, and Lee et al. (2024) show that large language models may introduce bias across the educational AI lifecycle. These concerns make learner protection central to any education-focused AI policy.

Lastly, the outer layer, *participation and implementation governance*, examines who participates in AI policy design, implementation, monitoring, and revision. This layer speaks directly to Diem et al.'s (2014) focus on power and voice by asking who sits at the decision-making table, who is absent, and whose knowledge counts. Radu (2021) shows that national AI policy processes often privilege government, industry, academia, and technical experts. In education, however, Arnstein's (1969) work on participation reminds us that affected groups should have meaningful influence over policy direction, while Gaber (2020) cautions against participation that merely legitimizes decisions already made. Teachers, learners, parents, school leaders, teacher educators, rural communities, disability advocates, language communities, and local communities should therefore be treated as policy actors, not only as policy recipients (Burriss et al., 2024). The critical policy analysis and the Education-Centered AI Policy Framework allowed us to examine both what Ghana's AI strategy says about education and what it leaves underdeveloped. Diem et al. (2014) provide the critical foundation for asking questions about power, voice, silences, assumptions, and policy-practice gaps. The Education-Centered AI Policy Framework translates those concerns into education-specific dimensions for analyzing national AI policy.

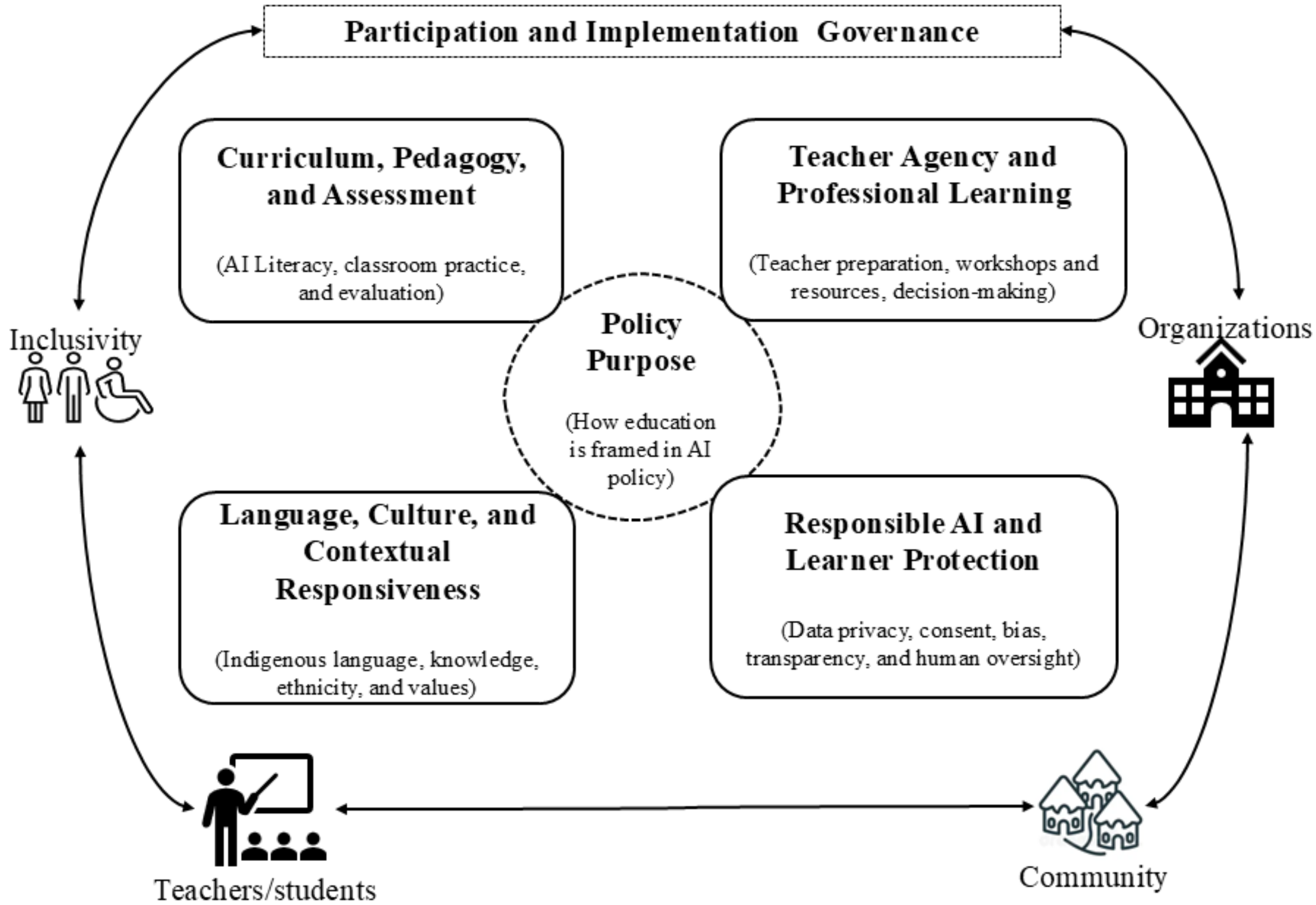

*Figure 1: Education-Centered AI Policy Framework*

## Researchers' Positionality

The research team includes Ghanaian education researchers, teacher educators, and AI-in-education scholars with experience in teacher preparation, educational technology, AI policy, rural education, and culturally responsive pedagogy. These backgrounds supported close attention to Ghanaian classroom realities, local language inclusion, teacher agency, and implementation concerns. At the same time, because several authors are professionally invested in AI and education in Ghana, the team used reflexive memoing, shared coding review, and repeated discussion of textual evidence to reduce interpretive overreach.

## Method

## Research Design

This study used a critical qualitative policy document analysis guided by the Education-Centered AI Policy Framework presented in Figure 1 (Bowen, 2009; Diem, et al., 2014). This approach was appropriate because the study examined how Ghana's *National Artificial Intelligence Strategy, 2025–2035* frames education, positions stakeholders, defines implementation priorities, and addresses responsible AI concerns within educational settings. Rather than treating the strategy as a neutral technical roadmap, the analysis approached it as a policy text that constructs meanings of education, AI readiness, inclusion, governance, and national development.

**Data Sources and Policy Context**

Following Bowen's (2009) approach to qualitative document analysis, this study examined Ghana's *National Artificial Intelligence Strategy, 2025-2035* as the primary policy document. The strategy was selected using the *Center for AI and Digital Policy's Significant AI Policy News* criteria because it is an official, publicly available, recent, and national-level AI policy document with clear implications for education and governance (Center for AI and Digital Policy, 2026). Developed under the Ministry of Communication, Digital Technology and Innovations, the strategy was publicly launched in Accra on April 24, 2026, by President John Dramani Mahama (MoCDTI, 2026), although the policy document itself is cited as MoCDTI (2025). The strategy identifies education as the first of eight pillars and positions AI as central to Ghana's goal of becoming an AI-enabled and innovation-driven society by 2035. For this study, the education-related sections were treated as the main units of analysis.

**Analytical Framework**

The analysis was guided by the Education-Centered AI Policy Framework developed in the previous section. As shown in Figure 1, the framework includes six connected components: *policy purpose; teacher agency and professional learning; curriculum, pedagogy, and assessment; language, culture, and contextual responsiveness; responsible AI and learner protection; and participation and implementation governance*. (see Appendix A) These components allowed us to examine the strategy's first pillar as an education policy document through at least five (5) critical policy questions for each component, with attention to its implications for teachers, learners, schools, communities, and education-sector governance. The coding framework was developed deductively from this framework and its grounding in critical policy analysis (Diem et al., 2014). The main coding categories were the six framework components, with document-level transparency and coherence added as a cross-cutting analytic code.

**Analytical Procedure**

Consistent with Bowen's (2009) approach to qualitative document analysis, the team analyzed Ghana's *National Artificial Intelligence Strategy, 2025-2035* through iterative skimming, close reading, and interpretation. The first reading provided a broad understanding of the document's structure, policy priorities, implementation language, and education-related claims. The team then extracted all education-related references into a shared coding matrix. The research team used the Education-Centered AI Policy Framework to code the document deductively across the six components. Because the framework is grounded in critical policy analysis, we examined how the strategy constructed education, positioned stakeholders, privileged particular priorities, and left some school-level concerns silent or underdeveloped (Diem et al., 2014).

Document-level transparency and coherence was also included as a cross-cutting analytic code because it connected directly to responsible AI, disclosure, policy communication, and internal consistency. To strengthen consistency, we met weekly over approximately six weeks to compare codes, discuss disagreements, refine definitions, and review whether interpretations were supported by textual evidence. For example, references to training “AI-ready youth” were coded under policy purpose because they reflected how the strategy framed education in relation to national AI readiness. They were also coded under curriculum, pedagogy, and assessment when they referred to AI literacy, coding, data science, or skills development. Such excerpts were interpreted as evidence of a workforce-oriented framing when they emphasized national skills development more than classroom implementation, assessment practice, teacher agency, or learner protection in our findings.

## Findings

Analysis of Ghana’s National Artificial Intelligence Strategy 2025-2035 revealed a policy vision that is ambitious, developmental, and future-oriented, but unevenly translated into education. The strategy positions education as central to Ghana’s AI transformation, yet its dominant emphasis is on talent production, employability, and economic competitiveness. Across the document, classroom-level questions concerning teacher agency, teacher preparation, pedagogy, assessment, learner protection, cultural responsiveness, local language instruction, and AI disclosure remain less developed. Six related findings emerged from the analysis.

### Document-Level Transparency and Coherence Concerns

Before examining the education-specific provisions, the strategy reveals two document-level concerns. First, the visual design appears to include AI-styled or AI-assisted content without visible disclosure (see Figure 2). This is important because the strategy itself recognizes that “the lack of transparency and explainability in some AI systems... makes it difficult to detect bias, audit and trust AI systems, and hold them accountable” (MoCDTI, 2025, p. 26). In this sense, a national AI strategy that promotes Responsible AI should also model transparency in its own production and presentation. If AI-generated or AI-assisted visuals, icons, or design elements were used, they should be clearly acknowledged.

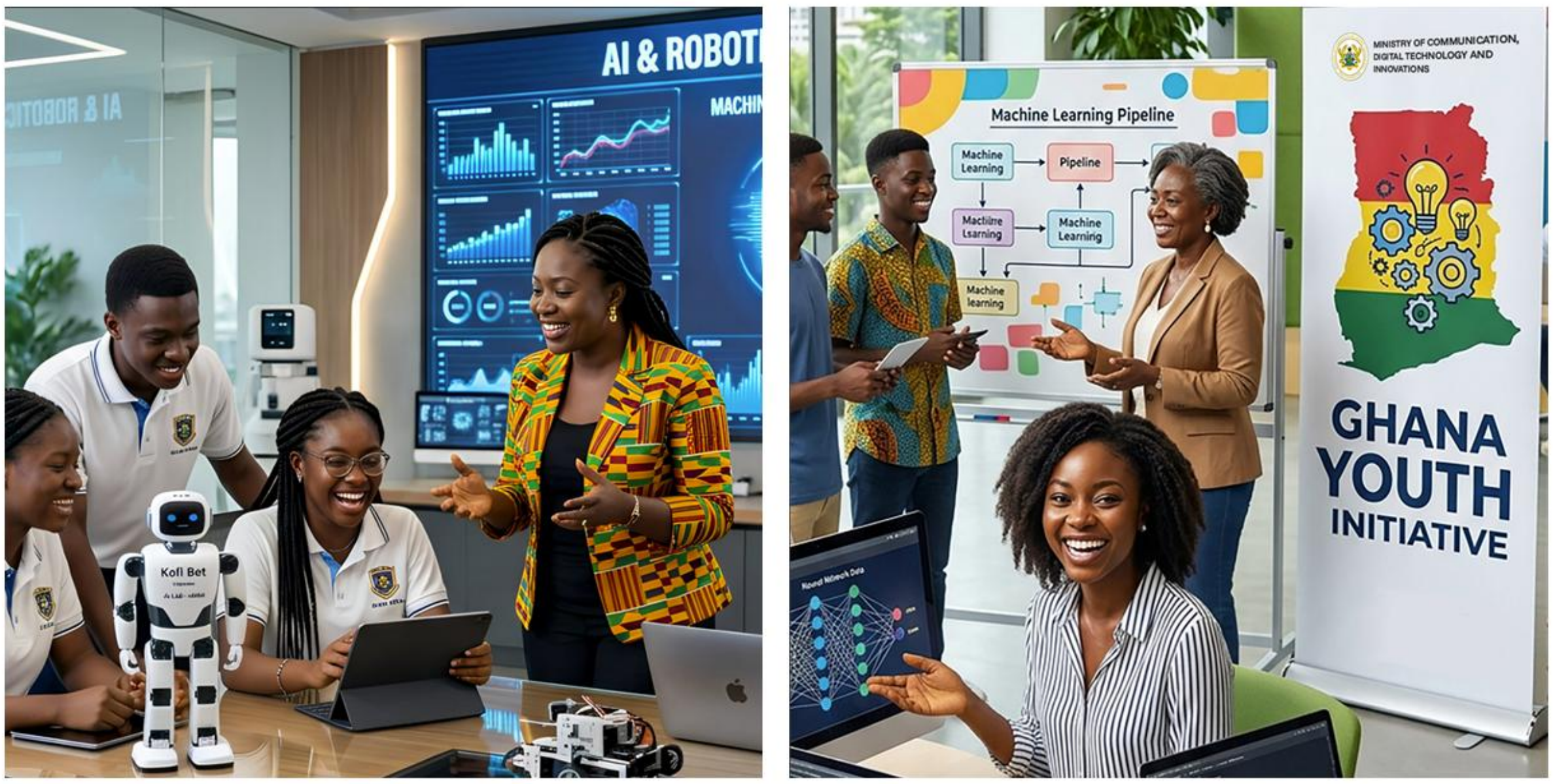

*Figure 2: Example of apparent AI-styled visual content without visible disclosure (Screenshot,* MoCDTI, 2025, pp. 38, 44).

Second, there appears to be a mismatch between the figure presenting the vision and mission statements and the way these statements are developed in the text (see Figure 3). This matters because policy coherence depends not only on ambition, but also on accuracy, clarity, and internal alignment. The strategy's own concern with "accurate and high quality data" and the risks of "less accurate outcomes" provides a useful basis for reading accuracy as a Responsible AI issue, not only a technical concern. Thus, the document-level mismatch between visual representation and textual explanation weakens the strategy's internal coherence.

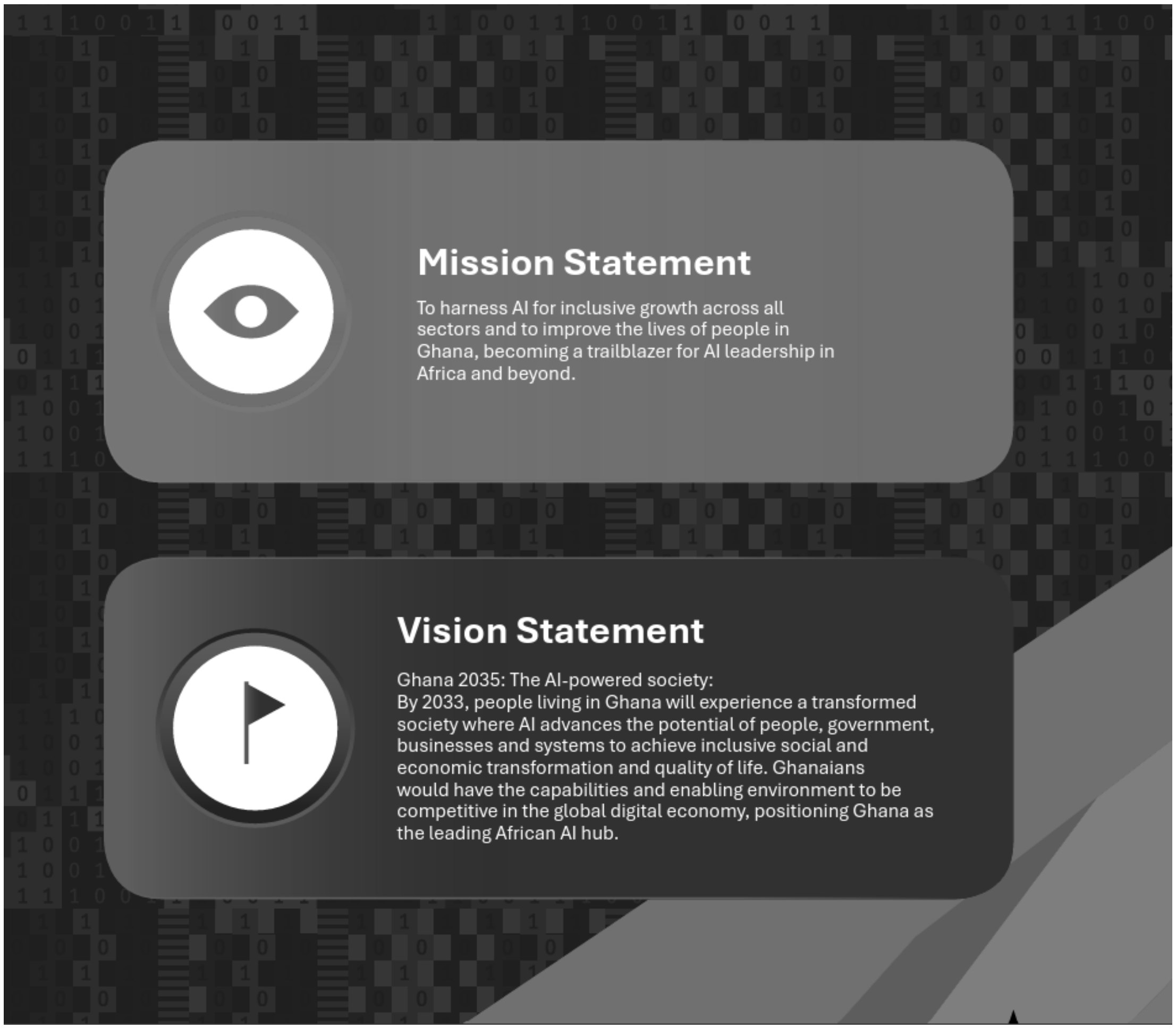

*Figure 3: Screenshot of mismatch between representative icon image and text (MoCDTI, 2025, p. 20).*

These concerns do not dismiss the value of the strategy. Rather, they show that Responsible AI must begin with the policy document itself. Clear disclosure of AI-assisted content and stronger alignment between figures and textual explanations would make the strategy more consistent, credible, and educationally useful as a public document before it is translated into schools, teacher education, curriculum, and public communication.

**Policy Purpose: Education as AI Workforce Readiness**

The strategy presents education primarily as a mechanism for preparing Ghana's population to participate in an AI-driven economy. Schools, universities, Technical Vocational Education and Training (TVET) institutions, training centres, and professional development programmes are positioned as part of a national talent pipeline. This workforce logic is captured clearly in the strategy's talent development pillar:

> In an era of rapid technological and economic changes, it is essential to equip people in Ghana with the skills to survive and thrive - now and in the future. This pillar develops

> AI talent by increasing the number of graduates with AI and machine learning, data science, data engineering, computer science, and practical technical skills (e.g. data preparation, data collection and labelling)... By increasing AI talent, individuals and the nation can participate in the global AI and digital economy (MoCDTI, 2025, pp. 29-30).

The repeated emphasis on "skills," "AI talent," "graduates," "technical skills," and the "global AI and digital economy" frames education as a route to national productivity and global competitiveness. This framing is reinforced by the strategy's call to "conduct a baseline study to identify the current needs and availability of talent with specific skills and competencies for AI in the labour market to inform curricula in schools and universities" (p. 38). Here, curriculum planning is explicitly tied to labour-market demand.

The strategy's most ambitious education target appears in the proposed AI Ready Ghana programme:

> Launch an Initiative that aims to train over 1,000,000 AI-ready youth by 2033. The Initiative would ensure AI training for youth in the first year of high school through their last years in tertiary education" (MoCDTI, 2025, p. 38).

This target demonstrates the seriousness with which the strategy treats AI literacy. However, the limitation is that education is mainly narrated through employability, productivity, and economic transition. Learners are imagined largely as future workers, entrepreneurs, innovators, and contributors to the digital economy. Less visible are broader educational purposes such as civic formation, cultural identity, critical consciousness, ethical reasoning, creativity, democratic participation, and social responsibility. Thus, while the strategy is strong on AI workforce readiness, it requires a fuller educational philosophy that balances economic participation with holistic human development.

**Stakeholder Representation**

The strategy presents AI development as a national multi-stakeholder agenda, but the actors most visible in the document are mainly government agencies, technical institutions, universities, entrepreneurs, and industry actors (see Figure 4). The acknowledgements foreground institutional and technical stakeholders:

> The Government of Ghana deeply appreciates the Ministry of Communication, Digital Technology and Innovations (MoCDTI) for its strategic leadership. We also acknowledge the crucial coordination of the Data Protection Commission (DPC). We thank key Ministries, Departments, and Agencies (MDAs), including the National Information Technology Agency (NITA), Cyber Security Authority, National Communications Authority (NCA), and Ministry of Environment, Science, and Technology (MEST), for their important policy inputs" (MoCDTI, 2025, p. 16).

This representation shows strong national coordination, but education-facing stakeholders are less visible in the formal policy conversation. Teachers, students, parents, school leaders, teacher unions, parent-teacher associations, Colleges of Education, and basic school communities are not strongly positioned as policy actors. This matters because the strategy expects education to carry much of Ghana's AI transformation, yet those closest to classroom implementation appear mainly as beneficiaries or recipients of training rather than contributors to policy design.

## Key Stakeholders

Given many initiatives and policies for digital development in Ghana, it is essential to coordinate and harmonise across all of the relevant stakeholders from the public sector (regulators, ministries, departments, and agencies), private sector (corporates and start-ups), civil society, and academia.

### Public Sector

Key public sector actors important to the National AI Strategy include:

- Ministry of Communications and Digitalisation (MoCD)
  - Cyber Security Authority
  - Data Protection Commission
  - Kofi Annan Centre
- National Communications Authority (NCA)
- Ministry of Environment, Science, Technology, and Innovation (MESTI)
- Ministry of Lands and Natural Resources (MLR)
- Ministry of Finance and Economic Planning (MoF/MoFEP)
- Bank of Ghana (BoG)
- National Information Technology Agency (NITA)
- National Development Planning Commission (NDPC)
- Ghana Revenue Authority
- Ministry of Gender

Ministries that can advise and support responsible AI adoption in key sectors, such as the Ministry of Food and Agriculture, the Ministry of Health, the Ministry of Trade

*Figure 4: Screenshot of key stakeholders from policy document(MoCDTI, 2025, Appendix)*

This limited representation creates an implementation concern. AI in education will affect curriculum, assessment, learner data, teacher workload, academic integrity, multilingual instruction, and classroom equity. Therefore, the strategy would be stronger if it positioned teachers, learners, parents, teacher educators, and school communities not only as beneficiaries of AI policy, but as active stakeholders in shaping responsible AI education.

**Teacher Preparation, Curriculum, and Assessment**

The strategy identifies curriculum reform as a major pathway for preparing Ghanaian learners for the AI age. It calls for coding, AI skills, data science, data protection, and data ethics to be introduced across schooling:

> Update existing STEM and IT curricula in secondary education to incorporate practical coding and AI skills, basics of data ethics, data protection and data science, and raise awareness about jobs in AI and digital fields. Prepare students by incorporating the basics of data science and coding in primary education. This can include teaching with resources like MIT's Scratch and using no-code tools. Students should also be taught on how to leverage existing AI across different use cases to improve their productivity (MoCDTI, 2025, p. 38).

This passage is important because it expands AI education beyond programming alone. It includes data ethics, data protection, productivity, and early exposure to coding and data science. However, it names curricular content without providing a classroom roadmap. It does not specify age-appropriate learning progressions for primary, junior high, senior high, TVET, or teacher education. It also does not show how teachers should scaffold AI concepts, adapt AI tools for diverse learners, or assess AI-supported learning. This gap becomes more visible in how teachers are positioned. The strategy recommends that Ghana should "promote training courses for teachers" and work "with Ghana Education Services (GES) ICT coordinators to train teachers" (p. 39). It further suggests existing programmes such as "the Smart Africa Digital Academy (SADA) and AIMS' online hybrid course for teachers in mathematics" (p. 39). These recommendations acknowledge the need for professional learning, but they frame teachers mainly as recipients of training rather than as professional agents who interpret curriculum, design learning, protect students, and make ethical instructional decisions.

For teacher education, this is a major gap. The strategy does not provide a detailed AI teacher competency framework. It does not explain what teachers should know about AI literacy, prompt use, AI bias, data protection, academic integrity, assessment redesign, learner privacy, inclusive pedagogy, and responsible classroom use. It also gives limited attention to Colleges of Education, teacher education universities, pre-service teacher preparation, and continuous professional development systems. Yet these institutions are central to preparing teachers who will enact AI policy in classrooms. The same under-specification appears in assessment. The strategy encourages students to learn "how to leverage existing AI across different use cases to improve their productivity" (p. 38), but it does not provide guidance on how such AI-supported work should be disclosed, assessed, or graded. This omission is significant because AI use directly affects homework, essays, projects, examinations, originality, authorship, plagiarism, and academic integrity. The strategy states that AI Ready Ghana should "include training on AI ethics in general as a core part of the programme" and should:

> Expand beyond data protection and cybersecurity to cover other ethical dimensions outlined in the UNESCO Recommendation on the Ethics of AI, including environmental sustainability, cultural protection, fairness, inclusion and the future of work (MoCDTI, 2025, p. 39).

Although this ethical framing is valuable, it remains broad. It does not translate into school-level assessment rules for AI disclosure, acceptable AI assistance, teacher feedback, student authorship, or responsible use during examinations. The strategy therefore provides a strong national curriculum ambition, but leaves teachers without the detailed preparation, pedagogical guidance, and assessment rules needed for everyday classroom implementation.

**Culture and Language**

The strategy recognizes local languages as important to Ghana's AI future. It proposes to "use local languages to structure datasets" and to "fund initiatives to collect, transcribe, and label datasets in various local languages" (MoCDTI, 2025, pp. 52). This is important because Ghana is multilingual, and AI systems that ignore Ghanaian languages may reproduce exclusion. Local language datasets can help AI tools become more useful, inclusive, and nationally relevant. However, the strategy treats language mainly as a resource for AI system development rather than as a resource for classroom teaching and learning. There is a difference between collecting Ghanaian language data for AI models and using Ghanaian languages to support understanding, participation, and culturally responsive pedagogy in schools. The strategy does not clearly explain how AI could support bilingual instruction, concept explanation, classroom dialogue, or learner participation in languages such as Twi, Ewe, Ga, Dagbani, Fante, Nzema, and others.

This is a missed educational opportunity. In Ghanaian classrooms, language affects comprehension, confidence, identity, and inclusion. If AI is to support education, it should not only process local languages as datasets. It should also help teachers explain difficult concepts in learners' familiar languages, connect instruction to Ghanaian cultural contexts, and support students who struggle with English-only instruction. The culture gap is similar. The strategy mentions ethical dimensions such as "cultural protection" (p. 39), but it does not develop a clear classroom-level account of culturally responsive AI education. It does not explain how AI-supported instruction can draw on Ghanaian histories, community knowledge, indigenous practices, local examples, or learners' cultural identities. Thus, culture and language are recognized at the level of national AI development, but they are not sufficiently translated into multilingual and culturally responsive classroom practice.

**Inclusion and Classroom Practice**

The strategy gives serious attention to inclusion, particularly for rural youth, women, persons with disabilities, informal sector workers, and underserved communities. It proposes that government and partners should…"deliver AI and digital skills training through rural tech hubs, mobile labs, and local TVET centres" (MoCDTI, 2025, p. 44).

The same section recommends scholarships, stipends, free access to AI learning materials, regional AI learning centres, mentorship, local media, radio, and "community outreach in local languages" (p. 44). These proposals show that the strategy does not imagine AI education only for elite schools or urban universities. It recognizes that access must be expanded through flexible, community-based, and regionally distributed models. The strategy also addresses disability and gender inclusion. It recommends that Ghana should "develop specialized AI training programs tailored for individuals with disabilities and informal sector workers to enhance their employability" and "create accessibility resources and tools that enable participation in AI education for disabled individuals and informal sector participants" (p. 39). It also proposes a Women in AI initiative with "scholarships, mentorship, and venture support" and sets "a 40% gender participation target for AI fellowships and technical training programs" (p. 40).

However, inclusion is framed mainly as access to programmes, training, centres, and opportunities. Less attention is given to how inclusion will occur inside ordinary classrooms. The strategy does not sufficiently explain how teachers will support learners with special educational needs, how rural girls will be protected from cost and connectivity barriers, how gender stereotypes in technology will be addressed pedagogically, or how school leaders will sustain AI inclusion in under-resourced settings. Infrastructure further complicates the inclusion agenda. The strategy acknowledges that "internet penetration is incomplete, particularly in rural areas, and poses challenges in terms of affordability and reliability" (p. 32). It also reports that 4G penetration is 41% in rural areas compared with 88% in urban centres (pp. 27, 32). These figures show that AI education will not be implemented on equal ground. Without sustained attention to electricity, devices, connectivity, maintenance, teacher support, and safe learning spaces, AI education may benefit urban and well-resourced schools more than rural and deprived schools. Thus, the strategy names inclusion well, but it needs a stronger account of inclusive classroom practice and rural implementation.

**Responsible AI Governance and Learner Protection**

The strategy places Responsible AI at the centre of Ghana's national AI agenda. It proposes that the government will "establish an independent, well-resourced Responsible AI Authority (RAI Authority)" to coordinate, monitor, and guide ethical AI development (MoCDTI, 2026, pp. 6, 15). It also identifies risks such as "algorithmic bias" and "misuse of personal data" and considers "mandatory insurance requirements for high-risk AI-generated outputs to protect individuals' rights and wellbeing" (pp. 6, 25, 57). These statements show that the strategy recognizes AI as a social, ethical, and regulatory issue, not only a technical innovation agenda.

However, the responsible AI framework is stronger at the national level than at the school level. The strategy does not clearly address learner profiling, student data commercialization, AI surveillance in schools, parental consent, child consent, automated grading, behavioural monitoring, or the use of AI systems to track learners. This is a serious educational gap because children's educational data are sensitive, and AI systems can shape learning opportunities, teacher judgments, feedback, and school decisions.

This concern also connects back to disclosure. The strategy recommends that Ghana should "provide training for media personnel to strengthen responsible AI communication and public awareness" (p. 40). However, responsible AI communication should also be demonstrated in official policy documents and educational materials. If AI-generated or AI-assisted content is used in public communication, it should be labelled. Otherwise, the policy risks teaching transparency in principle while leaving opacity in practice.

These findings show that Ghana's AI strategy is ambitious and timely, particularly in its commitment to AI literacy, youth training, TVET, internships, rural outreach, gender inclusion, disability inclusion, local language data, and responsible AI governance. However, the strategy is stronger on national ambition and workforce planning than on school-level educational implementation. It prepares Ghana's youth for the AI economy, but it says less about teacher agency, pre-service teacher education, classroom pedagogy, assessment policy, AI disclosure, multilingual instruction, culturally responsive pedagogy, child-centred learner protection, and sustainable rural implementation. For the strategy to become educationally transformative, it must move beyond workforce readiness and develop a more education-specific implementation framework that connects national AI goals to the daily realities of Ghanaian teachers, learners, parents, school leaders, and communities.

**Discussion**

The findings show that Ghana's *National Artificial Intelligence Strategy, 2025-2035* frames education primarily through national AI readiness, skills development, and workforce preparation. This is evident in the strategy's emphasis on training "AI-ready youth," expanding TVET, developing AI talent, promoting data science, and aligning curricula with labour-market needs (MoCDTI, 2025). This finding reflects Schiff's (2022) distinction between "education for AI" and "AI for education." Ghana's strategy is strong on education for AI because it prepares learners to participate in the digital economy. Shi (2025) similarly shows that many national AI strategies prioritize reskilling, higher education reform, workforce development, and national competitiveness. However, the Ghana case shows the limits of this dominant policy logic.

The strategy identifies education as the first pillar, which signals its importance, but the document gives less attention to how AI will be enacted in ordinary classrooms. It names coding, AI skills, data science, data ethics, and data protection, but it does not provide clear learning progressions for primary, junior high, senior high, TVET, or teacher education (MoCDTI, 2025). This gap matters because AI literacy is not simply technical exposure. For example, Touretzky et al. (2019) argue that AI learning should be age-appropriate, while Casal-Otero et al. (2023) show the need for coherent K-12 AI literacy models. Chee et al. (2025) further frames AI literacy as involving data literacy, algorithmic understanding, ethics, communication, problem solving, and critical judgment. The implication is that Ghana's strategy has the right ambition, but it needs a stronger education-sector translation that clarifies what learners should know, and how AI learning should progress across levels and disciplines, and what teachers should teach.

The findings further show that although the strategy recommends teacher courses and collaboration with Ghana Education Service ICT coordinators, it says less about teachers as curriculum interpreters, assessment designers, ethical decision-makers, and protectors of learners (MoCDTI, 2025). This is a major policy gap because AI integration depends heavily on teacher judgment. For instance, Mishra et al. (2023) argue that AI requires expanded technological, pedagogical, and contextual knowledge, and therefore teacher judgment becomes paramount. In addition to this, Zhai (2024) similarly positions teacher agency as moving away from observing AI to adopting, collaborating with, and innovating through AI. Again, studies such as that of Vaughan and Wah (2026) and Yadav et al. (2025) also show that pre-service teachers need preparation in ethics, inclusion, facilitation, and critical evaluation, not only AI tool use. For Ghana, this means Colleges of Education, teacher education universities, the Ghana Education Service, and the National Teaching Council should be central to AI policy implementation. Without a national teacher AI competency pathway, the strategy risks building national AI ambition without preparing the professionals who must translate that ambition into classroom practice.

The findings on curriculum, assessment, language, and culture show that school-level guidance remains underdeveloped. Even though the strategy encourages students to use AI for productivity, it does not explain how AI-supported work should be disclosed, assessed, graded, or treated in examinations. This silence is significant in Ghana's examination-driven system (Anapey, 2026), where unclear AI assessment rules could lead to confusion, uneven enforcement, and mistrust. Perkins et al. (2024) argue that institutions need explicit levels of acceptable AI use, while Xia et al. (2024) show that AI is reshaping feedback, task design, academic integrity, transparency, and human judgment. The language finding raises a related concern. Ghana's strategy rightly supports local language datasets, but language is treated mainly as data for AI systems rather than as a pedagogical resource for teaching and learning. Owu-Ewie and Eshun (2019) show that Ghanaian classrooms are multilingual spaces where language shapes comprehension, participation, and identity. Nyaaba and Zhai (2025) and Nyaaba et al. (2025) further show that culturally responsive AI becomes more meaningful when it includes local languages, indigenous knowledge, curriculum-aligned examples, and culturally familiar contexts. Thus, Ghana's AI policy should connect local language development to bilingual instruction, culturally responsive pedagogy, and classroom participation, not only to dataset creation.

The findings show that responsible AI must be modeled through learner protection, disclosure, and participatory governance. Ghana's strategy gives important attention to Responsible AI governance, rural outreach, gender inclusion, disability inclusion, and local language access (MoCDTI, 2025). However, it provides less school-level guidance on learner profiling, child consent, parental consent, AI surveillance, student data commercialization, automated grading, appeal processes, and human oversight. Nguyen et al. (2022) argues that responsible AI in education requires transparency, privacy, fairness, accountability, and human-centered governance. Holstein and Doroudi (2021) warn that AI can reproduce educational inequalities, and Lee et al. (2024) show that large language models may introduce bias across the educational AI lifecycle. The document-level finding on apparent AI-styled visual content without visible disclosure strengthens this point, if official policy documents expect responsible AI use from schools and citizens, they should also model transparency in their own communication.

Lastly, the analysis showed that education-related stakeholder engagement did not include extensive representation of teachers, students, parents, guardians, communities, and families, even though these groups are central to school-level AI implementation. Participation is therefore not a secondary issue but an implementation condition. Radu (2021) shows that national AI policy processes often privilege government, industry, academia, and technical experts. Arnstein (1969) and Gaber (2020) further remind us that participation is weak when affected groups only receive policy rather than shape it. For Ghana, teachers, learners, parents, school leaders, Colleges of Education, rural communities, disability advocates, and language communities should therefore be treated as policy actors. Without their voices, Ghana's AI strategy may remain technically ambitious but weakly connected to classroom realities, learner needs, and community trust.

**Implications and recommendations**

The findings point to the need for a sector-specific, education-centered AI implementation framework for Ghana. This recommendation aligns with the strategy's broader call for AI adoption across sectors but argues that education requires its own framework because schools involve children, teachers, assessment systems, local languages, families, communities, and unequal infrastructure. Such a framework should preserve the strategy's strong workforce agenda while translating national AI ambition into school-level guidance for teachers, learners, school leaders, teacher education institutions, parents, and communities.

A key implication is that the development of this framework should be participatory. Teachers, learners, parents, guardians, school leaders, Colleges of Education, teacher education universities, rural communities, disability advocates, language communities, curriculum bodies, and education researchers should help shape the framework (Radu, 2021). Additionally, participation should include consultations, classroom-based research, regional dialogues, and studies across basic education, secondary education, TVET, and teacher education. Another implication concerns curriculum and assessment. Ghana should rethink AI literacy as a progressive educational agenda developed across primary, junior high, senior high, TVET, higher education, and teacher education (Nyaaba, 2025). This curriculum should include data literacy, algorithmic understanding, ethical reasoning, creativity, communication, problem solving, critical judgment, and responsible use. Schools also need clear assessment guidance on AI disclosure, authorship, acceptable assistance, grading, examinations, originality, feedback, and human review (Nyaaba, 2025).

Moreover, Ghana should develop a national teacher AI competency framework for pre-service and in-service teachers, anchored in Colleges of Education, teacher education universities, the Ghana Education Service, the National Teaching Council, and continuous professional development systems (Nyaaba, 2025; Zhai, 2024). This framework should prepare teachers to evaluate AI outputs, detect bias, protect learner data, redesign assessment, use AI responsibly across subjects, support multilingual pedagogy, and adapt AI tools to Ghanaian classrooms (UNESCO, 2021). Teachers should be treated as professional agents in AI implementation, not only as recipients of training. Furthermore, rural and under-resourced schools need reliable electricity, affordable internet, devices, maintenance support, and safe digital learning spaces. Ghana should also explore offline or low-bandwidth AI tools where internet access is weak (Nyaaba et al., 2026). Without this infrastructure layer, AI education may widen the digital divide by benefiting urban and well-resourced schools more than rural and deprived communities.

The framework should also support responsible partnerships for locally responsive AI tools. Collaboration with frontier AI companies, local EdTech developers, Ghanaian researchers, teacher educators, curriculum experts, and cultural and language specialists can support the design of Ghana-centered and regionally responsive AI tools. For example, Nyaaba, Kyeremeh, et al. (2026) developed a conversational AI agent for teacher education in Ghana, showing how AI tools can be designed for local educational needs. Such tools should reflect local languages, indigenous knowledge, curriculum expectations, and culturally meaningful examples. This is important because Nyaaba, Wright, et al. (2026) show that generative AI can reproduce digital neocolonialism through Western curriculum ideologies, language marginalization, cultural imperialism, racial and ethnic underrepresentation, pedagogical control, and access inequity. Locally responsive AI development can therefore reduce dependence on externally designed systems and support more equitable, culturally grounded AI use in Ghanaian education. These recommendations suggest that Ghana's next step is not simply to implement AI in education, but to design an education-centered AI policy pathway. Such a pathway should connect workforce readiness with teacher preparation, curriculum reform, assessment redesign, learner protection, infrastructure, local language instruction, culturally responsive pedagogy, locally responsive AI tools, and participatory governance.

## Limitations

This study has some limitations. First, the analysis was based on the publicly available online version of Ghana's National Artificial Intelligence Strategy 2025-2035. We recognize that there may be related preparatory documents, implementation plans, consultation reports, or sector-specific materials that were not publicly accessible at the time of analysis. Access to such documents could have provided additional context and possibly a more detailed understanding of the education-related provisions. Second, the study focused mainly on the education-related dimensions of the strategy, especially the pillars that speak most directly to AI talent development, youth readiness, training, and responsible AI implementation. Although this focus was appropriate for the purpose of the study, a broader analysis of all policy sectors could reveal additional connections to education. Lastly, some of the gaps identified in this study may be addressed in future implementation documents. However, because the national strategy is the main public-facing policy document, we argue that clearer guidance on teacher preparation, curriculum, assessment, learner protection, language, culture, and school-level implementation would strengthen the strategy itself and make it more useful for education stakeholders.

## Conclusion

Although national AI strategies are increasingly analyzed for their attention to governance, workforce development, innovation, and competitiveness, less is known about how these strategies frame education as a sector with its own pedagogical, cultural, ethical, and implementation demands. To address this gap, we developed and applied an *Education-Centered AI Policy Framework* to examine Ghana's *National Artificial Intelligence Strategy, 2025–2035* through an education-sector lens. Using critical qualitative policy document analysis, we analyzed the strategy deductively through the key components of the framework: *policy purpose, teacher agency, curriculum and assessment, language and culture, learner protection, and participatory governance*. The framework helped us move beyond asking whether education appears in the strategy to examine how education is framed, whose voices are visible, what educational priorities are emphasized, and what school-level concerns remain underdeveloped.

Our analysis of Ghana's strategy shows that the document is ambitious, timely, and nationally important. It gives strong attention to AI literacy, youth skills, TVET, workforce readiness, rural outreach, inclusion, local language data, and responsible AI governance. However, we also found that the education agenda is stronger on national AI readiness than on school-level implementation. Issues such as teacher agency, pre-service teacher education, curriculum progression, assessment guidance, AI disclosure, child-centered learner protection, multilingual pedagogy, culturally responsive AI use, and participatory governance remain underdeveloped.

These findings matter for Ghana because AI education cannot be reduced to training youth for the digital economy. It also requires rethinking what students should learn, how teachers should be prepared, how assessment should change, how parents and communities should be involved, and how learners should be protected. Ghana needs clearer AI curriculum pathways across basic education, secondary education, TVET, higher education, and teacher education. It also needs professional development systems that prepare teachers to evaluate AI outputs, detect bias, protect learner data, support multilingual pedagogy, and use AI responsibly across subjects. The findings also point to the need for Ghana-specific and locally responsive AI tools. We therefore argue that Ghana's next step is not simply to implement AI in education, but to design a sector-specific, education-centered AI implementation pathway.

DECLARATION OF GENERATIVE AI SOFTWARE TOOLS IN THE WRITING PROCESS

During the preparation of this work, the authors used ChatGPT/Claude and Grammarly to support grammar correction, language refinement, and overall clarity of expression. These tools were used only to improve readability and presentation, not to generate new data, conduct analysis, or make substantive changes to the interpretation of the findings. After using these tools, the authors reviewed and edited all content and took full responsibility for the content of the publication.

**Appendix A**

*Coding Framework*

The analysis was guided by the Education-Centered AI Policy Framework, grounded in critical policy analysis. Following Diem et al. (2014), the coding examined how Ghana's AI strategy constructs education, whose voices are visible, what assumptions shape the policy, what silences remain, and how the policy may affect schools, teachers, and learners.

| Main code | Critical policy questions | Supporting sources |
|---|---|---|
| Policy purpose | 1. How is education framed in the AI strategy?<br>2. Is education presented mainly as workforce preparation, human capital development, social transformation, or democratic participation?<br>3. What assumptions does the strategy make about the role of AI in national development?<br>4. Does the strategy balance economic competitiveness with broader educational aims such as equity, inclusion, ethics, and citizenship?<br>5. What educational purposes are emphasized, minimized, or left silent? | Schiff, 2022; Shi, 2025 |
| Participation and implementation governance | 1. Who is visible in the design, implementation, and governance of the AI strategy?<br>2. Are teachers, learners, parents, school leaders, teacher educators, and rural communities included as policy actors?<br>3. Which institutions are given authority to implement or govern AI in education?<br>4. Does the strategy provide mechanisms for consultation, feedback, accountability, and local participation?<br>5. Whose voices appear to be privileged, and whose voices are absent or marginalized? | Diem et al., 2014; Radu, 2021; Arnstein, 1969; Gaber, 2020 |
| Teacher agency and professional learning | 1. Are teachers positioned as professional agents or mainly as recipients of AI training?<br>2. Does the strategy recognize teachers' role in curriculum interpretation, assessment design, ethical judgment, and learner protection?<br>3. Does the strategy provide guidance for pre-service teacher education and in-service professional development?<br>4. Are teacher educators, Colleges of Education, and teacher education universities included in AI capacity-building plans?<br>5. Does the strategy prepare teachers to evaluate AI outputs, detect bias, adapt AI tools, and use AI responsibly in context? | Mishra et al., 2023; Zhai, 2024; Yadav et al., 2025 |
| Curriculum, pedagogy, and assessment | 1. Does the strategy provide clear guidance on AI literacy across levels of education?<br>2. Are curriculum pathways specified for basic education, secondary education, TVET, higher education, and teacher education?<br>3. Does the strategy address how AI should be used in teaching and learning across subjects?<br>4. Does the strategy provide guidance on assessment, academic integrity, authorship, grading, disclosure, and acceptable AI assistance?<br>5. Are pedagogical issues such as creativity, critical thinking, problem solving, and human judgment addressed? | Touretzky et al., 2019; Casal-Otero et al., 2023; Chee et al., 2025; Perkins et al., 2024; Xia et al., 2024 |
| Language, culture, and contextual responsiveness | 1. Are Ghanaian languages, cultural knowledge, and local educational realities addressed in the strategy?<br>2. Does the strategy treat local languages only as data resources, or also as pedagogical resources for teaching and learning? | Owu-Ewie & Eshun, 2019; Issaka et al., 2024; Wiafe et al., 2025; Nyaaba & Zhai, 2025 |
| | 3. Are rural schools, under-resourced schools, and multilingual classrooms considered in implementation plans?<br>4. Does the strategy support culturally responsive AI tools, examples, and curriculum materials?<br>5. Does the strategy address how AI may reproduce cultural bias, language marginalization, or dependence on externally designed systems? | |
| Responsible AI and learner protection | 1. Does the strategy provide concrete safeguards for learners in AI-supported education?<br>2. Are issues of student data privacy, child consent, parental consent, surveillance, and profiling addressed?<br>3. Does the strategy provide guidance on automated grading, algorithmic decision-making, human oversight, and appeal processes?<br>4. Are fairness, transparency, accountability, and bias mitigation translated into school-level guidance?<br>5. Does the strategy protect vulnerable learners, including children with disabilities, rural learners, and learners from low-resource contexts? | Nguyen et al., 2022; Holstein & Doroudi, 2021; Lee et al., 2024 |
| Document-level transparency and coherence | 1. Does the policy document model the responsible AI practices it promotes?<br>2. Is there visible disclosure where AI-generated or AI-assisted content appears to have been used?<br>3. Are figures, visuals, and textual explanations internally consistent?<br>4. Are key terms such as AI literacy, responsible AI, inclusion, and education clearly defined and used consistently?<br>5. Does the document show coherence between its vision, mission, pillars, implementation plans, and education-related recommendations? | Nguyen et al., 2022; Perkins et al., 2024 |